# Four-dimensional Vibrational Spectroscopy for Nanoscale Mapping of Phonon Dispersion in BN Nanotubes

Ruishi Qi[1,2#], Ning Li[1,2#], Jinlong Du[1], Ruochen Shi[1,2], Yang Huang[3,4], Xiaoxia Yang[5], Lei Liu[6], Zhi Xu[7], Qing Dai[5], Dapeng Yu[8], and Peng Gao[1,2,9*]

**Direct measurement of local phonon dispersion in individual nanostructures can greatly advance our understanding of their electrical, thermal, and mechanical properties. However, such experimental measurements require extremely high detection sensitivity and combined spatial, energy and momentum resolutions, thus has been elusive. Here, we develop a four-dimensional electron energy loss spectroscopy (4D-EELS) technique based a monochromated scanning transmission electron microscope (STEM)[1-3], and present the position-dependent phonon dispersion measurement in individual boron nitride nanotubes (BNNTs). Our measurement shows that the unfolded phonon dispersion of multi-walled BNNTs is close to hexagonal-boron nitride (h-BN) crystals, suggesting that interlayer coupling and curved geometry have no substantial impacts on phonon dispersion. We also find that the acoustic phonons are extremely sensitive to momentum-dependent defect scattering, while optical phonons are much less susceptible. This work not only provides useful insights into vibrational properties of BNNTs, but also demonstrates huge prospects of the developed 4D-EELS technique in nanoscale phonon dispersion measurements.**

1 Electron Microscopy Laboratory, School of Physics, Peking University, Beijing 100871, China.
2 International Center for Quantum Materials, Peking University, Beijing 100871, China.
3 School of Materials Science and Engineering, Hebei University of Technology, Tianjin 300130, China.
4 Hebei Key Laboratory of Boron Nitride Micro and Nano Materials, Hebei University of Technology, Tianjin 300130, China.
5 Division of Nanophotonics, CAS Center for Excellence in Nanoscience, National Center for Nanoscience and Technology, Beijing 100190, China.
6 Department of Materials Science and Engineering, College of Engineering, Peking University, Beijing 100871, China.
7 Songshan Lake Materials Lab, Institute of Physics, Chinese Academy of Sciences, Guangdong, China.
8 Shenzhen Key Laboratory of Quantum Science and Engineering, Shenzhen 518055, China.
9 Collaborative Innovation Center of Quantum Matter, and Beijing Key Laboratory of Quantum Devices, Beijing 100871, China.
# These authors contribute equally to this work.
* Corresponding author. E-mail: p-gao@pku.edu.cn.

Phonon plays a fundamental role in mechanical, electrical and thermal properties of materials. There has been significant interest in measuring phonon dispersion in the hope of gaining mechanistic understanding and optimization of materials' properties in materials science and condensed matter physics. However, such a measurement is very challenging for crystal defects and nanostructures, where the tiny size requires high spatial resolution and high detection sensitivity. Although the tip enhanced Raman spectroscopy and scanning near-field optical microscopy[4] can reach reasonable spatial resolution up to a few tens of nanometers, their momentum transfer is much smaller than the typical Brillouin zone (BZ) size and thus inaccessible to the high momentum phonons. Other vibrational spectroscopy techniques such as inelastic X-ray/neutron scattering are capable of measuring phonon dispersions for bulk crystals, but the lack of spatial resolution (limited by their beam size and low sensitivity) results in scattering signal averaged over large crystals or ensembles of nanostructures, hence precluding phonon dispersion measurements in individual nanostructures[5,6].

Recent developments of aberration correctors and monochromators within in STEMs have enabled kiloelectronvolt electron beams with sub-10 meV energy resolution and atomic spatial resolution, extending EELS measurements into lattice vibration properties in the past decade[1,7]. Many spatially resolved measurements such as atomic-resolved phonon spectroscopy[8-11], phonon polariton (PhP) mapping[3,12], isotope identification[13] and temperature measurement[14] are now attainable. Recently, momentum-resolved vibrational measurements of h-BN and graphite flakes using STEM-EELS have also been reported[2,15]. Although previous studies have demonstrated high sensitivity and large momentum transfer of this technique, they only focused on flakes and two-dimensional (2D) sheets that are essentially homogeneous in space, and thus didn't take advantage of the high spatial resolution of electron microscopes. Long acquisition time (up to ~10 h for each dispersion diagram at a single spatial location) of the serial acquisition method[2] also precludes the possibility to perform a 2D scan or even a line scan. To date, nanoscale position-dependent phonon dispersion measurement in a single nanostructure has not been reported to the best of our knowledge.

Here, we take a step further by using momentum-resolved EELS (M-EELS) to efficiently measure the phonon dispersion at different positions in a single nanostructure. By using a slot aperture, phonon dispersion data can be acquired on parallel within much shorter acquisition time[16]. This allows us to add two spatial dimensions to this technique and enables 4D vibrational EELS measurements in a single nanostructure at nanometer scale. As a model system, BNNTs have shown remarkable mechanical, thermal and nano-optical properties that promise a wide range of applications[17-20]. These properties are dominated by lattice vibrations because the electronic contributions are nearly negligible due to a large bandgap. Therefore, the knowledge about their phonon properties are highly desired to advance our understandings. *Ab initio* and zone-folding calculations have predicted an unfolded phonon dispersion of BNNTs similar to that of two dimensional h-BN layers[21], but experimental verification has been absent. In addition, in multi-walled BNNTs interlayer coupling usually lead to vacancies, distortions and change of stacking order[17,22-24], whose impacts on phonon properties still await experimental exploration. Unfortunately, traditional spectroscopy techniques cannot measure their phonon dispersion even for ensembles of BNNTs, not to mention for individual ones. This is because their cylindrical shape results in a crystal orientation varying with the azimuthal angle, which means broad-beam scattering scheme will inevitably mix up scattering signals with momenta along all directions; non-uniform chirality and diameter distribution among BNNTs will also mix up phonon signals from BNNTs with different structures and/or orientations. With fine spatial resolution substantially smaller than the tube diameter, the 4D-EELS technique is evidently ideal to probe their local vibration properties.

Our measurements reveal that the unfolded phonon dispersion of multiwalled BNNTs is very similar to that of h-BN crystals, which indicates geometry of atom planes and interlayer coupling have no substantial impact on the phonon dispersion. Real-space mapping of different phonon modes suggests that acoustic phonon modes in BNNTs are susceptible to defect scattering and can reveal defects that are nearly invisible in traditional STEM images, while high-frequency optical modes are less sensitive. Momentum dependence of defect scattering is also revealed. These results provide useful information about phonons and their associated properties of BNNTs. The 4D-EELS technique also opens up new possibilities to probe the defect modes of phonons in many systems such as thermoelectric materials, nanoelectronic devices, and interfacial superconductors.

The experimental setup is illustrated in Fig. 1a. A kiloelectronvolt electron beam is focused to nanometer-wide, with its spatial and energy resolutions further refined by the aberration correctors and monochromator. After balancing the spatial and momentum resolution, the typical beam size (spatial resolution) is estimated to be ~4 nm while keeping a momentum resolution of ~ 0.3 Å$^{-1}$ (Supplementary Fig. 1). After scattering from the sample, electrons pass through magnetic lenses to form diffraction spots on the diffraction plane, where an EELS aperture is placed to select scattered electrons with particular momentum. Previous M-EELS studies collect electrons by a small entrance aperture, which is moved relative to the diffraction plane to acquire phonon dispersion data serially[2,15]. A more efficient way for M-EELS data acquisition, which reduces the typical acquisition time for each dispersion diagram from ~10 h to tens of minutes, is to use a slot aperture to select a narrow line on the diffraction plane[25,26], producing a 2D intensity map versus both energy and momentum transfer. This scheme is also advantageous for data consistency among momentum points (avoiding beam instabilities over time) and for a much higher momentum sampling density. With the beam scanning in two spatial dimensions, a 4D dataset is recorded. The sample morphology is also recorded using high-angle scattered electrons at the same time, giving us information about the microstructure of BNNTs. Top panel of Fig. 1b is a high-angle annular dark field (HAADF) image of a typical BNNT, from which the inner and outer radius can be easily determined to be ~13 nm and ~51 nm. From the electron diffraction patterns shown in the bottom panel, its chirality can be determined close to zigzag. This allows us to characterize the structure of individual BNNTs under investigation.

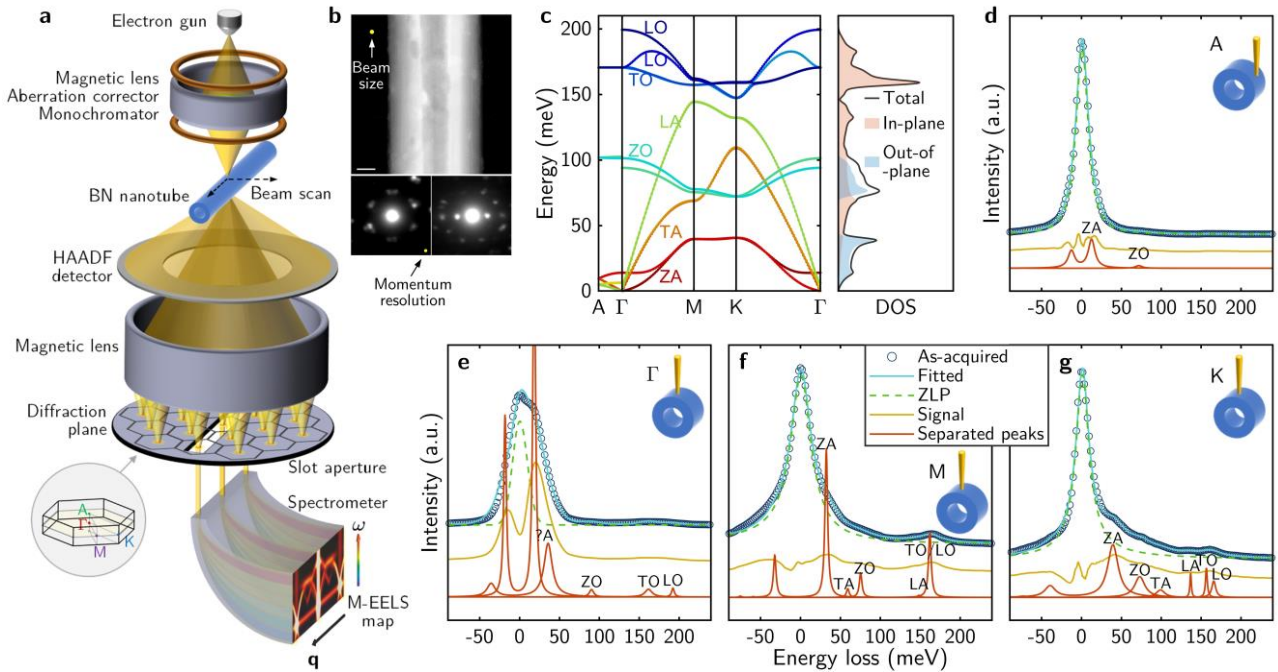

**Figure 1 | 4D vibrational EELS in a STEM. a**, Schematic of the experimental setup. The inset at the bottom shows high-symmetry points in the first BZ of h-BN. **b**, A typical HAADF image of a BNNT (top panel). Scale bar, 20 nm. Two bottom panels are electron diffraction patterns with the beam located at the center (left) and near the edge (right) of the tube. **c**, Phonon dispersion curve and phonon DOS of h-BN crystal calculated by DFPT. Light pink and light blue shadows are in-plane and out-of-plane contributions. **d-g**, Typical EEL spectra at high symmetry points A, Γ, M and K in the unfolded BZ of a BNNT. Dark-blue circles represent the raw spectra, to which solid cyan lines are fitted (Methods). Dashed green lines are quasi-elastic lines extracted from the fitting, which are subtracted from the measured spectra to get phonon signal (yellow lines). Red curves are separated multiple phonon peaks, the summation of which, once convoluted with the fitted quasi-elastic line, gives the least difference with the measured signal. For clarity lines are vertically shifted.

Fig. 1d-g show typical EEL spectra at high-symmetry points of the unfolded BZ, A, Γ, M and K respectively (see inset of Fig. 1a). Apart from a strong quasi-elastic line located at zero energy (zero loss peak, ZLP), in the range from ~10 meV to 200 meV we observe multiple phonon modes with their intensity and energy varying with the momentum transfer. Because of the worse energy resolution and large variation of phonon peak intensity and energy, removal of the quasi-elastic line (background subtraction) is not as straightforward as in traditional vibrational EELS. Here we develop a fitting method (Methods) to remove the quasi-elastic line (green dashed lines in Fig. 1d-g) and simultaneously separate phonon peaks (red traces). These phonon peaks can be assigned to six types (ZA, TA, LA, ZO, TO and LO) by comparing with the calculated phonon dispersion. Fig. 1c shows the density functional perturbation theory (DFPT) calculation of phonon dispersion and phonon density of states (DOS) for a bulk h-BN crystal. Phonon peaks in the experimental data matches reasonably well with the calculation.

To visualize spatial-dependent phonon dispersion in a single BNNT, we performed two 4D-EELS measurements for a zigzag BNNT. Supplementary Fig. 2 shows the HAADF image and electron diffraction patterns of the BNNT under investigation. Supplementary Video 1 and 2 show two 4D datasets acquired with the slot aperture placed in two directions schematically shown in Fig. 2a. As a concise summary, Fig. 2b and 2c show line profiles with the beam scanning from the tube center (top panels) to the edge (bottom panels), during which the local phonon dispersion substantially changes. The position-dependent dispersion diagrams can be well captured by viewing the nanotube locally as h-BN flakes rotated by a spatial-dependent tilt angle.

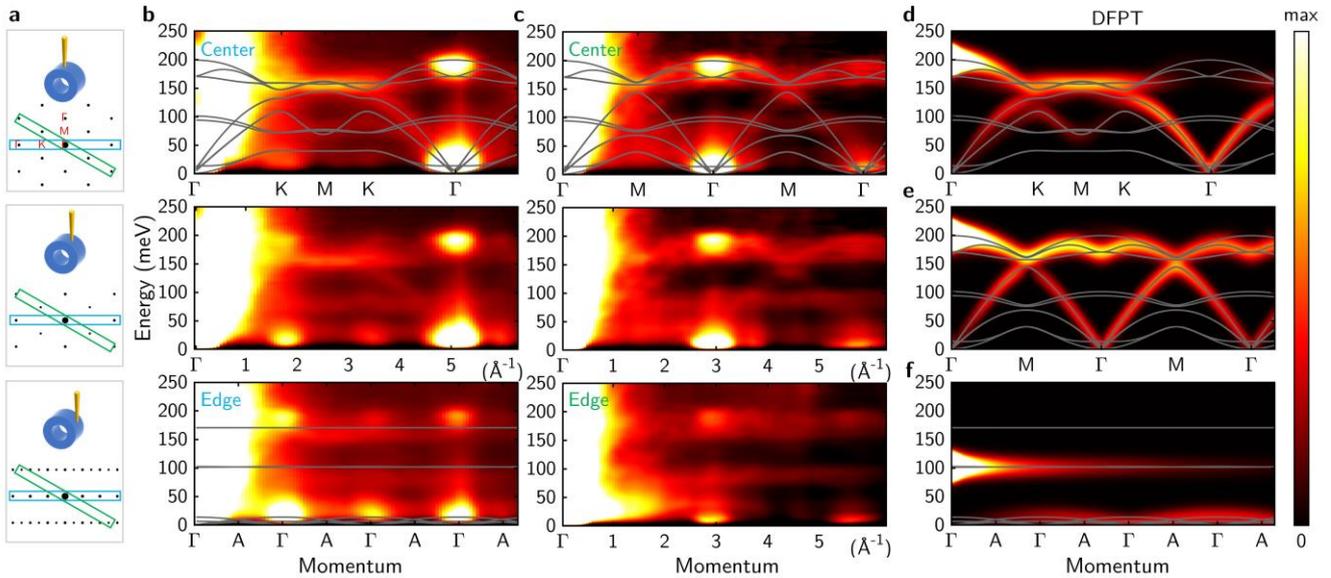

**Figure 2 | 4D-EELS measurement of phonon dispersion in an individual zigzag BNNT. a,** Schematic of the experimental geometry. Blue and green rectangles illustrate the position of the slot aperture used in **b** (Supplementary Video 1) and **c** (Supplementary Video 2), respectively. The underlying black dot patterns are simulated electron diffraction patterns, in which the vertical direction is parallel to the tube axis. **b, c,** Phonon dispersion line profiles along a radius of the BNNT, acquired with the slot aperture placed as shown in **a**. Top and bottom panels correspond to the tube center and the edge respectively, and the middle panels are acquired between them. **d-f,** Calculated EELS intensity (corrected for statistical factor) for an infinite h-BN crystal along high-symmetry lines ΓKMKΓ (**d**), ΓMΓMΓ (**e**) and ΓAΓAΓ (**f**). Gray curves are DFPT calculation for bulk h-BN crystals.

When the beam passes through the center of the BNNT, the configuration is similar to the case of two vertically stacked h-BN flakes, with their relative rotation angle determined by the chiral angle of the tube. For BNNTs with small chiral angle (close to zigzag), the two flakes are rotationally aligned, so high-symmetry points Γ, M and K are well-defined for scattered electrons. Measured phonon dispersion diagrams along ΓKMKΓ and ΓMΓMΓ lines are displayed in top panels of Fig. 2b and 2c, respectively. As a reference, we also performed measurements for a h-BN flake under similar experimental conditions, with the results shown in Supplementary Fig. 3. As expected, they share common overall features as a manifestation of their similar local structure.

Approaching the edge, the dispersion gradually flattens as shown in middle and bottom panels of Fig. 2b, c. At the tube edge, the geometry is similar to tilting a h-BN flake along an armchair direction to 90°. Therefore, in the bottom panel of Fig. 2b we get the dispersion along ΓAΓAΓ line, where the phonon dispersion is highly non-dispersive due to the weak van der Waals (vdW) interlayer coupling. To understand and corroborate these measured spectra, Fig. 2d-f display calculated EEL spectra for bulk h-BN crystals along high-symmetry lines based on DFPT and scattering cross section formalism[2,3,27] (Methods). Because the scattering cross section is proportional to the inner product between the momentum transfer $q$ and the displacement vector of the $k^{th}$ atom and $\lambda^{th}$ mode $e_\lambda(k,q)$, only modes with non-zero vibration displacement along $q$ will be active in EELS measurements[27]. For h-BN flakes with normal electron beam incidence, ZO and ZA modes are not active since they have displacement vectors normal to atom planes, which are always perpendicular to the in-plane momentum transfer $q$. Nice agreement is found between the measured h-BN dispersion diagram (Supplementary Fig. 3) and the calculated spectra. For the BNNT, we observe that the intensity of ZO and ZA modes increases from the tube center to the edge, due to an increasing inner product between $q$ and $e_\lambda(k,q)$; meanwhile, the signal of in-plane modes becomes weaker. This is in qualitative agreement with scattering cross section theory. However, the ZO and ZA modes still have a small but non-zero intensity at the tube center, and the intensity of the TO/LO mode also remains considerable at the tube edge. This may due to the finite probe size and the curved geometry of atom shells, which slightly alters the vibration direction of each atom and leads to non-zero value of $q \cdot e_\lambda(k,q)$.

By averaging among multiple BZs, we extract unfolded phonon dispersion of the BNNT and the h-BN flake from the measured spectra, as depicted in Fig. 3. They both match well with the DFPT calculation, and their difference is beyond our energy resolution. The similarity in phonon dispersion of h-BN crystals and BNNTs implies that effects of structural curvature and interlayer coupling on phonon dispersion are insignificant. First, phonon dispersion in vdW crystals is insensitive to the shape of the atom plane. In fact, the zone-folding method for numerical study of nanotubes simply constructs phonon dispersion (or band structure) of nanotubes from that of a corresponding sheet with a periodic boundary condition imposed[21,28]. The underlying assumption of this scheme is that the effect of shell curvature is negligible, and this is supported by our observation at least for BNNTs with a radius of several tens of nanometers. *Ab initio* calculations have also shown that this method works very well for single-walled BNNTs[21] and carbon nanotubes[29] with smaller radius. Second, interlayer coupling in BNNTs can be different from that of a bulk h-BN crystal. Due to inevitably incommensurate perimeters between neighboring shells, the chiral angle, stacking order and interlayer distance may vary[22,30,31]. Supplementary Fig. 4 compares calculated phonon dispersions for AA' and AB staking bulk crystals, from which we can conclude that their dispersions are very similar, and weak vdW interlayer coupling does not substantially affect the phonon dispersion.

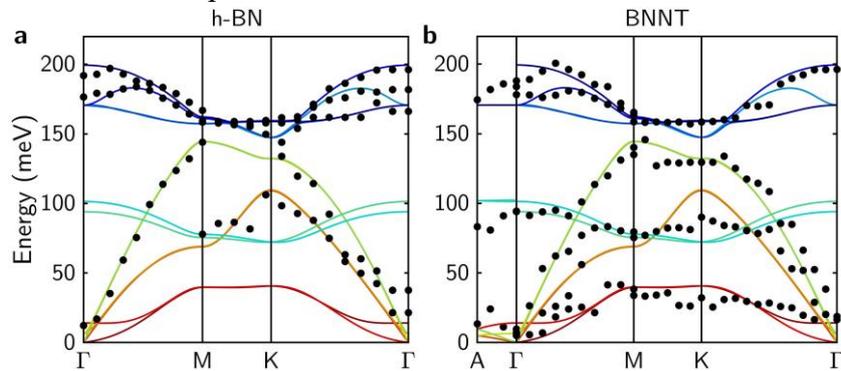

**Figure 3 | Phonon dispersion of the h-BN flake (a) and BNNT (b).** Measured dispersion is averaged among multiple BZs and shown as black dots. Solid curves are DFPT calculation for bulk h-BN crystals.

Note that due to large strain inherent in their structure, defects are common in multiwalled BNNTs[22]. With fine spatial resolution, we can actually map the intensity distribution of these phonon modes in real space and reveal the influence of defects. Shown in Fig. 4a is a HAADF image of a BNNT with obvious structural irregularities. On the top-left corner, the red ellipse marks a dark line in the image that extends far beyond the top boundary of this figure, which corresponds to a gap between atom shells in this BNNT. Red arrows highlight voids in some constituting shells. We select four momentum transfers (illustrated in Fig. 4b) to spatially map the phonon signal. Starting from the central diffraction spot, the momentum increases with a step size of ~2.5 Å$^{-1}$ perpendicular to the tube axis. At (nearly) zero momentum transfer, the electron beam interacts with lattice vibration of BNNTs predominately through dipole scattering, which essentially yields PhP signal[32]. In Fig. 4c, the PhP signal extends outside the tube (outlined by white dashed lines), and is not very sensitive to small defects. This is because the long-range Coulomb interaction is on a length scale comparable to the tube radius, which smears out small local variations. However, since the PhPs only have very small momentum compared with the BZ size, the phonon signal will be increasingly localized with larger momentum transfer. At the first BZ boundary (Fig. 4d) the spectra mainly contain impact scattering signal that is localized within the BNNT. The ZO mode is more intense near the edge, while TO/LO signal is stronger near the center, in agreement with the $\boldsymbol{q} \cdot \boldsymbol{e}_\lambda(k, \boldsymbol{q})$ dependence of the scattering cross section. At even higher momentum transfers, the quasi-elastic line is weaker so extracting acoustic phonon signal is possible. In Fig. 4e and 4f, the acoustic phonon maps show apparent correlation with the structral defects observed in the HAADF image, and the contrast is larger for higher momentum transfer. On the other hand, although optical branches are also slightly affected by the presence of voids, the contrast is much weaker.

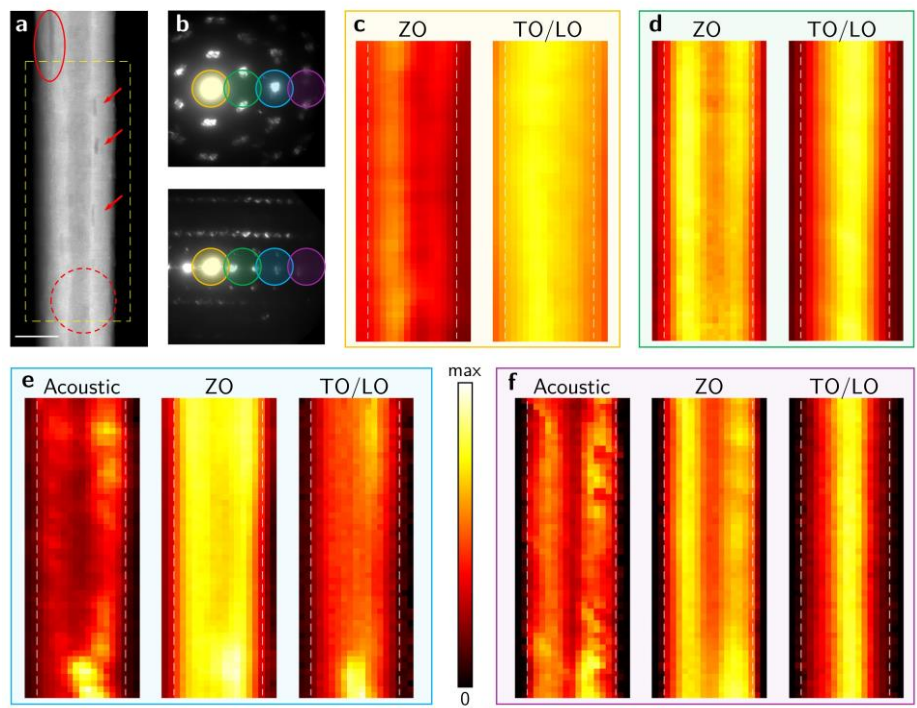

**Figure 4 | Real-space mapping of phonon signal at various momentum transfers. a**, HAADF image of a near-zigzag BNNT, in which structural defects can be clearly observed. Yellow dashed box encircles the scanning region in **c-f**. Scale bar, 50 nm. **b**, Electron diffraction patterns taken near the center (top panel) and the edge (bottom panel) of the BNNT. Colored circles denote four selected momentum transfers that correspond to **c-f**. The diffraction peak with highest intensity (encircled by the yellow circle) is the central diffraction spot with zero momentum transfer. The momentum transfer increases with a step of ~2.5 Å$^{-1}$ from **c** to **f**. **c-f**, Intensity maps of the acoustic, ZO and TO/LO phonon modes in real space. At small momentum transfers (**c** and **d**), the quasi-elastic line dominates the low-energy region, so the acoustic phonon signal is hard to extract.

Interestingly, a strong hot spot appears at the bottom these phonon intensity maps, while in this region no obvious contrast can be seen in the HAADF image (dashed circle in Fig. 4a). This defect scatters phonons most strongly at medium momentum transfers (Fig. 4e), and unlike voids, it also considerably scatters optical phonons. Being sensitive to defect scattering, acoustic phonon may be used to identify defects that are hard to visualize by traditional imaging techniques.

In summary, we have measured phonon dispersion relation and real-space phonon intensity distribution of individual BNNTs. We've shown that BNNTs have a similar phonon dispersion to that of bulk h-BN, which can be understood by the weak vdW interlayer coupling and supports the assumption underlying the commonly-used zone-folding calculation method. Real-space mapping emphasizes the importance of momentum-dependent defect scattering for understanding acoustic phonons. These conclusions should probably apply to many other similar materials such as carbon nanotubes. We have to admit, though, that more detailed measurements are still limited by the energy resolution. With higher resolution, the small change in dispersion associated with interlayer coupling may become detectable, then 4D-EELS could easily correlate phonon dispersion with small structural change such as chiral angle and local stacking order. Nevertheless, we anticipate that the demonstrated 4D-EELS technique may ultimately lead to the solution of phonon defect modes in solids[33] and will surely find more applications in vibrational measurements of numerous other interesting material systems.

**Acknowledgements**

The work was supported by the National Key R&D Program of China (grant no. 2016YFA0300903), the National Natural Science Foundation of China (11974023, 51672007), Key-Area Research and Development Program of Guangdong Province (2018B030327001, 2018B010109009), the National Equipment Program of China (ZDYZ2015-1), and the "2011 Program" from the Peking-Tsinghua-IOP Collaborative Innovation Center of Quantum Matter. We acknowledge Electron Microscopy Laboratory of Peking University for the use of electron microscopes and financial support. We thank Dr. Chenglong Shi and Dr. Tracy Lovejoy for helpful discussion.

**Author contributions**

R.Q. and N.L. contributed equally to this work. P.G., N.L. and R.Q. conceived the project. Y.H., Z.X. and L.L synthesized the BNNTs. N.L. acquired the 4D-EELS data. R.Q. processed the experimental data and performed DFPT calculations, with help and discussion from R.S., N.L. and J.D. R.Q and P.G. finalized the manuscript. P.G. supervised the project. All authors contributed to this work through useful discussion and/or comments to the manuscript.

**Competing interests**

The authors declare no competing interests.

## Methods

**EELS data acquisition**

The EELS data was acquired on a Nion U-HERMES200 microscope equipped with both a monochromator and aberration correctors, with 60 keV beam energy and ~12 pA beam current. Each 4D dataset was acquired with a total integration time of 36 min. To be more specific, the 2D raster scan was performed in a 100 nm × 150 nm region with 12 × 18 pixels, and the acquisition time of each pixel was 10 s. A small beam convergence semi-angle of 1.5 mrad was used for a satisfying momentum resolution, and the collection angle was 4.5 mrad × 75 mrad. A dense spectral sampling of 0.5 meV per channel was used. Under such conditions, the typical spatial, momentum and energy resolutions are 4 nm (estimated by comparing the acquired HAADF image to an ideal one), 0.3 Å$^{-1}$ (estimated by the diffraction spot size) and 20-30 meV, respectively (see Supplementary Fig. 1). All EELS data was acquired using Nion Swift software.

**EELS data processing**

For each 4D-EELS dataset, energy-momentum diagrams among different spatial pixels were registered by their 2D normalized cross correlation to correct possible beam energy drift. Spectra in different momenta are then aligned by their 1D cross correlation. After the alignment, we applied block-matching and 3D filtering (BM3D) algorithm to remove Gaussian noise[34,35], in which noise level is estimated based on high-frequency elements in the Fourier domain.

In vibrational EELS, the quasi-elastic line forms a strong ZLP in the measured spectra, the removal of which is usually not straightforward. In the study of surface phonon polaritons or LO/TO phonons, the signal of interest is usually of the order of 100 meV, which can be easily separated from the ZLP tail by fitting the background to a smooth fitting function (e.g., power function, third-order exponential, pseudo-Voigt function or Pearson-VII function[3,7,36]) in two energy windows, one before and one after the phonon peak (Supplementary Fig. 5a). However, for M-EELS data the energy loss peaks are at different energy for different momentum, so it's impossible to find a common energy window for all spectra. Besides, acoustic phonons at very low energy cannot be separated from the rapidly-varying background. In this work, the quasi-elastic line is removed by least-square fitting the measured spectrum (over the whole energy range, not only in some energy windows) to the following fitting function

$$P'(\omega) + \sum_\lambda a_\lambda \left[ L\left(\frac{\omega - \omega_\lambda}{c_\lambda}\right) + \exp\left(-\frac{\hbar\omega_\lambda}{k_B T}\right) L\left(\frac{\omega + \omega_\lambda}{c_\lambda}\right) \right] * P'(\omega) \qquad (1)$$

where the first term is a modified Pearson-VII function to account for the quasi-elastic line, and the second term is the contribution of multiple phonon modes in which $L(x)$ is the Lorentzian function. The asterisk symbol denotes convolution. The original Pearson-VII function is generally a Lorentzian raised by power $m$

$$P(\omega) = I_{\max} \frac{\gamma^{2m}}{\left[\gamma^2 + (2^{1/m} - 1)(2\omega - 2\omega_0)^2\right]^m}$$

in which $I_{\max}$, $\omega_0$ and $\gamma$ are the height, center and width of the Pearson peak. A modification (denoted by the prime symbol in Eq.1) is then made to account for possible ZLP asymmetry, where the constant width parameter $\gamma$ is replaced by a function $\gamma(\omega) = \frac{2\gamma_0}{1 + \exp(s\omega)}$ containing an additional fitting parameter $s$. With these 5 fitting parameters, the ZLP can be fitted rather well (Supplementary Fig. 5b). Assuming detailed balance, each active phonon mode contributes one energy gain peak and one energy loss peak with their intensity ratio being the Boltzmann factor $\exp -\frac{\hbar\omega_\lambda}{k_B T}$. The inelastic scattering signal are modeled by paired Lorentzian peaks with central energy $\pm\omega_\lambda$, peak height $a_\lambda$ and $a_\lambda \exp -\frac{\hbar\omega_\lambda}{k_B T}$, and peak width $c_\lambda$. These Lorentzian peaks are convoluted with the quasi-elastic line to account for the electron energy distribution. Each measured spectrum is fitted to the sum of these two contributions (Eq.1), and then the fitted ZLP is subtracted from the measured spectrum to get the signal. In the fitting, the calculated phonon energy was used as the initial guess. For most measured spectra, the fitting converges to the same minimum within a reasonable range of initial guess. A small portion of mis-identified peaks are manually removed. After removing the fitted ZLP, a correction for the statistical factor is then performed following literature[37].

**DFPT calculation**

The DFPT calculation was performed using local density approximation (LDA) and norm-conserving pseudopotentials with a 770 eV energy cutoff. The DOS was calculated based on phonon frequencies on a 23×23×8 mesh in the first BZ.

The scattering cross-section for an infinite bulk crystal is[2,3,27]

$$\frac{d^2\sigma}{d\omega d\Omega} \propto \sum_{\text{mode } \lambda} |F_\lambda(\boldsymbol{q})|^2 \left[ \frac{n_q+1}{\omega_\lambda(\boldsymbol{q})} \delta(\omega - \omega_\lambda(\boldsymbol{q})) + \frac{n_q}{\omega_\lambda(\boldsymbol{q})} \delta(\omega + \omega_\lambda(\boldsymbol{q})) \right]$$

where $\omega_\lambda(\boldsymbol{q})$ and $n_q$ are the frequency and occupancy number of the $\lambda^{\text{th}}$ phonon mode with wavevector $\boldsymbol{q}$. The two terms in the square brackets correspond to phonon emission and absorption process respectively. The coupling factor

$$F_\lambda(\boldsymbol{q}) \propto \frac{1}{q^2} \sum_{\text{atom } k} \frac{1}{\sqrt{M_k}} e^{-i\boldsymbol{q} \cdot \boldsymbol{r}_k} e^{-W_k(\boldsymbol{q})} Z_k(\boldsymbol{q}) \left[ \boldsymbol{e}_\lambda(k, \boldsymbol{q}) \cdot \boldsymbol{q} \right]$$

is determined by the mass $M_k$, real-space position $\boldsymbol{r}_k$, effective charge $Z_k(\boldsymbol{q})$, Debye-Waller factor[38] $\exp(-2W_k(\boldsymbol{q}))$ and phonon displacement vector $\boldsymbol{e}_\lambda(k, \boldsymbol{q})$ of $k^{\text{th}}$ atom in a unit cell. The effective charge $Z_k(\boldsymbol{q})$ was calculated using Eq.(9) in literature[27] with atomic form factors of B and N constructed from parameters in literature[39]. After correction for the statistical factor, the experimental diagram is essentially proportional to $\sum_\lambda |F_\lambda(\boldsymbol{q})|^2$, so the DFPT calculated diagrams in the main text also show this quantity (convoluted with a Gaussian to broaden the delta function for better visibility).

# Supplementary Figures

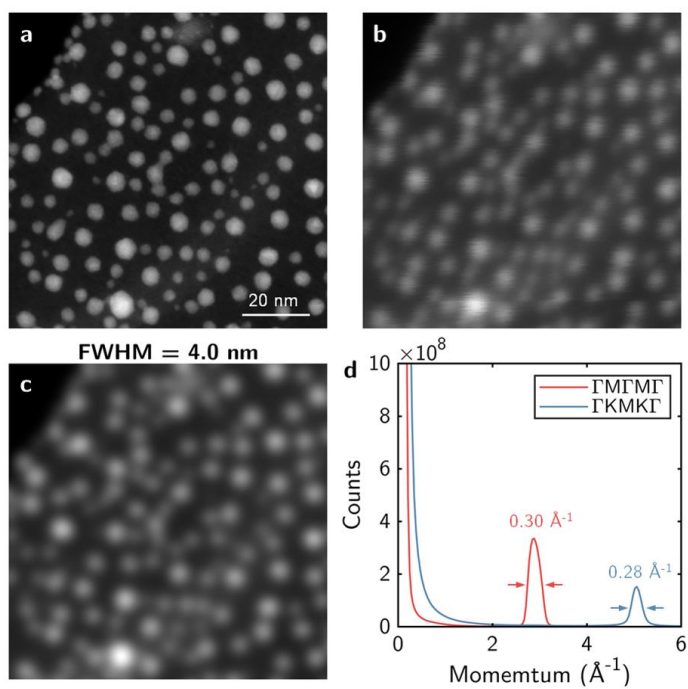

**Supplementary Figure 1 | Resolution estimation at 1.5 mrad convergence semi-angle. a**, HAADF image of gold nanoparticles taken with 20 mrad convergence semi-angle. The spatial resolution under such condition is known to be ~0.2 nm, so this image can serve as a reference. **b**, HAADF image taken with 1.5 mrad convergence semi-angle (used in our 4D-EELS measurements). **c**, The 20 mrad image convoluted with a Gaussian kernel, whose width is least-square fitted to achieve the best match between the convoluted image and the 1.5 mrad image. The fitted value of its FWHM is 4.0 nm. **d**, Momentum resolution estimation. Solid curves show total EELS counts (summed over all energy channels) as a function of the momentum transfer. Red and blue curves are acquired on a high-quality h-BN flake along ΓMΓMΓ and ΓKMKΓ lines, respectively. The FWHM of the Bragg reflection spots can be used to estimate our momentum resolution, which is ~0.3 Å$^{-1}$ for both datasets.

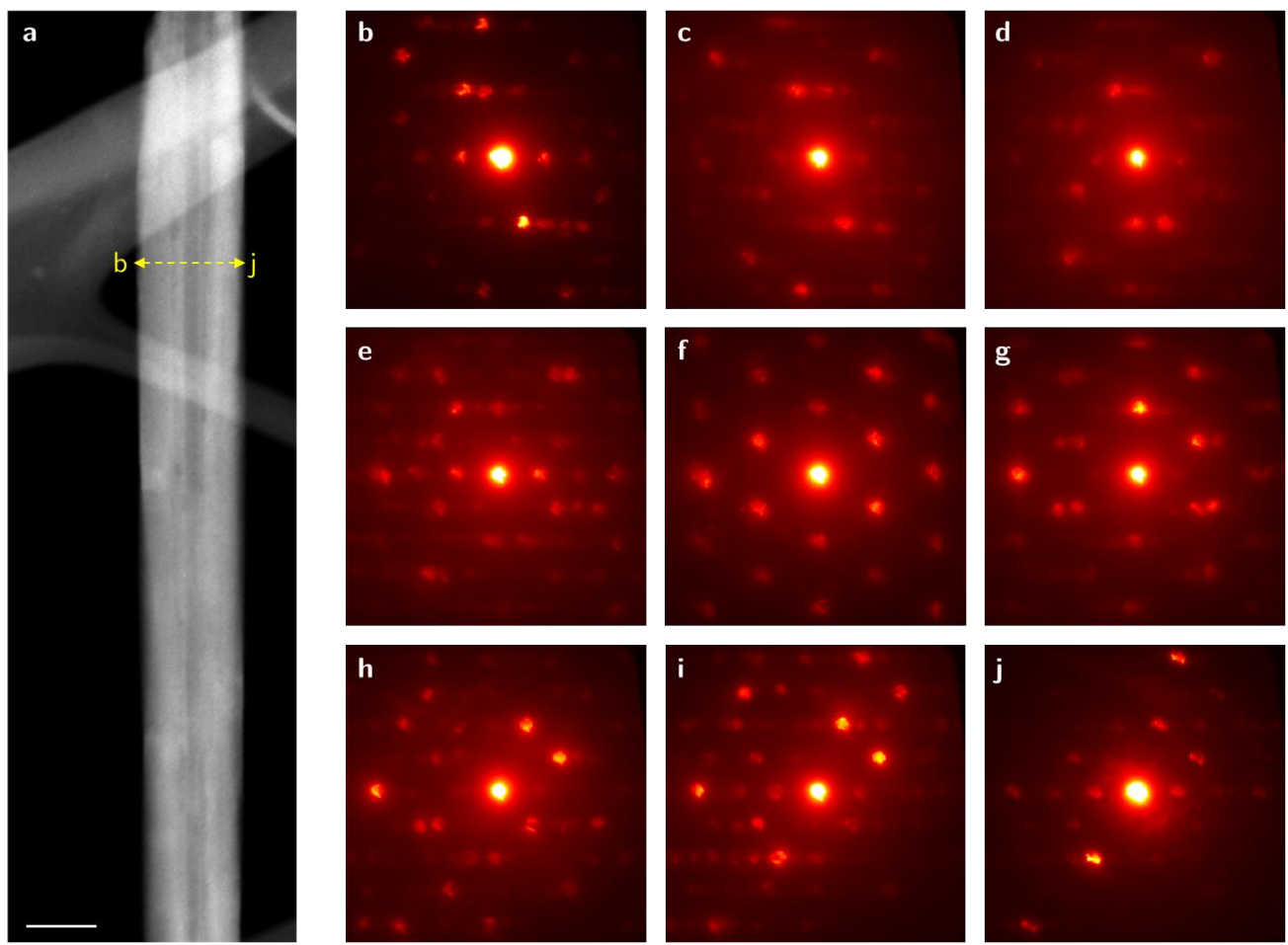

**Supplementary Figure 2 |** HAADF image (**a**) and electron diffraction patterns (**b-j**) across a diameter of the BNNT (yellow dashed arrow) under investigation in Fig. 2. Scale bar in **a**, 100 nm.

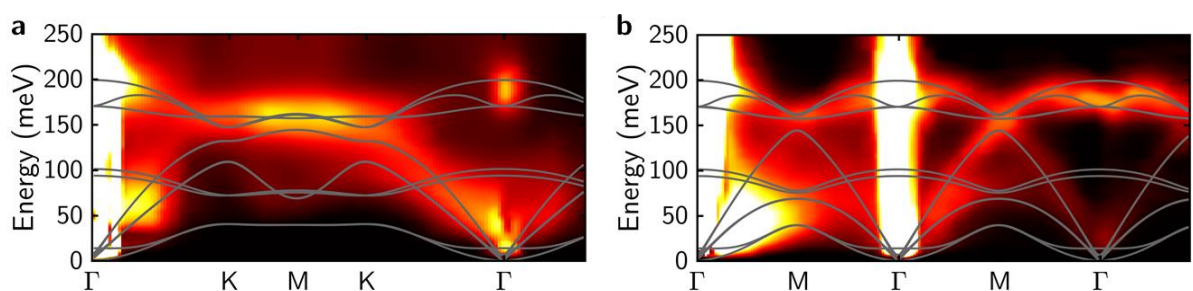

**Supplementary Figure 3 |** Measured phonon dispersion diagram of a h-BN flake along high-symmetry lines ΓKMKΓ (**a**) and ΓMΓMΓ (**b**). Solid curves are DFPT calculation for bulk h-BN crystals.

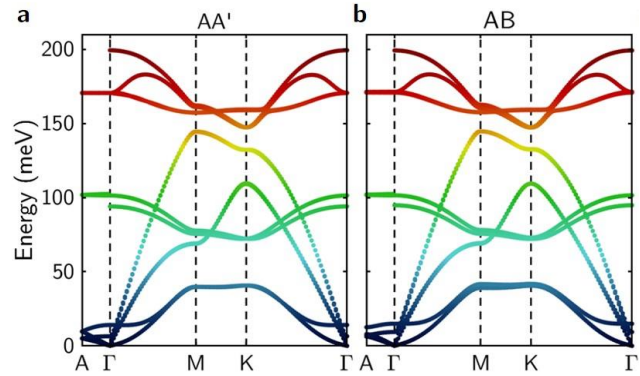

**Supplementary Figure 4 |** Calculated phonon dispersion of bulk BN crystals with AA' (**a**) and AB (**b**) stacking orders. Calculation parameters are listed in Methods.

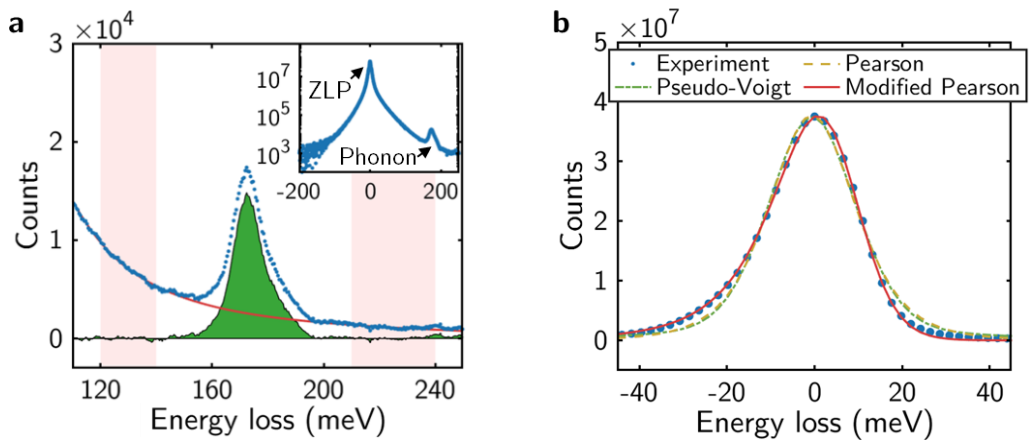

**Supplementary Figure 5 | Quasi-elastic line removal. a**, Double-window background subtraction for EELS data without momentum resolution (60 kV beam energy, 15 mrad convergence semi-angle, 8 meV energy resolution). Blue dots are as-acquired EELS data. Red trace, third-order exponential function fitted to the measured spectrum in two light pink windows. Green shadow, background-subtracted signal. Inset shows the overall spectrum in log scale. **b**, Quasi-elastic line modeling for momentum resolved EELS. Blue dots, typical spectrum acquired in a vacuum region far from any sample, containing the quasi-elastic line only. The energy resolution is 24 meV for this spectrum. Green and yellow curves are Pearson-VII function and pseudo-Voigt function fitted to the experimental data respectively, which deviate from the measured spectrum near ±20 meV asymmetrically. Red, modified Pearson function fitted to the spectrum.